\documentclass[floats,floatfix,showpacs,amssymb,prd,twocolumn,superscriptaddress,nofootinbib,reprint]{revtex4-2}

\usepackage{amssymb,amsmath,verbatim,mathtools,needspace,enumitem,etoolbox,graphicx,physics,microtype,afterpage,xspace,tabularx,lmodern,multirow}
\usepackage{siunitx}
\usepackage{gensymb}
\usepackage[dvipsnames, usenames]{xcolor}
\definecolor{linkcolor}{rgb}{0.0,0.3,0.5}
\usepackage[unicode, colorlinks=true, linkcolor=linkcolor, citecolor=linkcolor, filecolor=linkcolor, urlcolor=linkcolor, linktocpage, breaklinks]{hyperref}
\usepackage[all]{hypcap}
\usepackage[T1]{fontenc}
\usepackage[utf8]{inputenc}
\hypersetup{colorlinks=true,citecolor=romared,linkcolor=romared,urlcolor=romared}

\definecolor{romared}{RGB}{142,0,28}

\newcommand{\be}{\begin{equation}}
\newcommand{\ee}{\end{equation}}

\def\be{\begin{equation}}
\def\ee{\end{equation}}
\newcommand{\beq}{\begin{eqnarray}}
\newcommand{\eeq}{\end{eqnarray}}

\newcommand{\jcap}{JCAP}
\newcommand{\mnras}{Mon.\ Not.\ R. Astron.\ Soc.}

\newcommand{\bfB}{{\bf B}}
\newcommand{\bfx}{{\bf x}}
\newcommand{\bfk}{{\bf k}}

\newcommand{\hateps}{{\bf{\hat \epsilon}}}
\usepackage{orcidlink}

\begin{document}
\title{Primordial Density Perturbations from Magnetic Fields}

\author{Tal Adi\,\orcidlink{0000-0002-5763-9353}}
\email{talabadi@post.bgu.ac.il}
\affiliation{Department of Physics, Ben-Gurion University of the Negev, Be’er Sheva 84105, Israel}

\author{Hector Afonso G. Cruz\,\orcidlink{0000-0002-1775-3602}}
\email{hcruz2@jhu.edu}
 \affiliation{William H. Miller III Department of Physics and Astronomy, Johns Hopkins University, 3400 N. Charles Street, Baltimore, Maryland, 21218, USA}

 \author{Marc Kamionkowski\,\orcidlink{0000-0001-7018-2055}}
\email{kamion@jhu.edu}
 \affiliation{William H. Miller III Department of Physics and Astronomy, Johns Hopkins University, 3400 N. Charles Street, Baltimore, Maryland, 21218, USA}

\begin{abstract}
Perturbations to the cosmic baryon density---and thus to the total-matter density---can be induced by magnetohydronamic forces if there are primordial magnetic fields.  The power spectrum for these density perturbations was first provided in 1996, but without much in the way of detail in the derivation, and there has been confusion in the intervening years about this calculation.  In this brief note, we re-derive this power spectrum using modern conventions, provide a simplified result, and identify some of the discrepancies in the literature.
\end{abstract}

\date{\today}
\maketitle

%\tableofcontents

%\clearpage

\section{Introduction}

Although no current observations or measurements indicate the existence of magnetic fields in the early Universe, there is an abundance of models for new physics that predict primordial magnetic fields \cite{widrow02,subramanian16}.  In 1978, Ref.~\cite{wasserman78} showed that if there are primordial magnetic fields (PMFs), then magnetohydrodynamic effects induce density perturbations in the cosmic baryon density, and thus in the total-matter density.  The power spectrum for these density perturbations, for a given spectrum of magnetic fields, was then calculated in Ref.~\cite{kim96}.  This result was then reproduced in subsequent papers \cite{tashiro06a,gopal03,fedeli12,pandey13, err_shibusawa14, err_camera14} exploring various empirical consequences, and the results also used for numerical work in yet other papers~\cite{pandey15,Sanati:2020oay,Sethi:2004pe,Pandey:2012kk, Varalakshmi:2016sxx, Cheera:2018ofg}.

In this brief note, we re-derive the original result \cite{kim96} for the magnetic-field power spectrum using modern conventions and provide calculational details left out of the original work.  We also derive a simpler (and more easily evaluated) expression, and identify errors in some of the intervening literature.

\section{Equations of motion}

We surmise a primordial magnetic field $\bfB(\bfx,t)$, as a function of comoving position $\bfx$ and time $t$.  Prior to recombination, the tight coupling of the baryons to photons, as well as the large photon energy density, suppresses the dynamical effects of the magnetic fields.  After recombination, the baryons experience a magnetohydrodynamic force that provides a source to the linearized equation of motion for the fractional total-matter-density perturbation  $\delta_m(\bfx,t)$.  Following Refs.~\cite{wang20,kim96,Adi:2023qdf}, this equation is
\begin{equation}
     \ddot \delta_m + 2 H \dot \delta_m - 4 \pi G \bar \rho_m \delta_m = f_b \frac{v(\bfx)}{\mu_0 \bar \rho_{b,0} \left[a(t) \right]^3},
\label{eqn:EOM}     
\end{equation}
where $f_b=\bar\rho_b/\bar\rho_m$ is the fraction of the mean total-matter density $\bar\rho_m$ contributed by baryons; $\bar \rho_{b,0}$ the mean baryon density today; $a(t)$ the scale factor (normalized to unity today); $H=\dot{a}/a$ is the Hubble parameter; and $t_0$ the cosmic time today.  We recognize this as the usual equation for matter perturbations with a source, with
\begin{equation}
    v(\bfx)  \equiv \nabla \cdot \left\{ \bfB(\bfx) \times \left[\nabla \times \bfB(\bfx)\right] \right\}, 
\end{equation}
where we invoke the shorthand $\bfB(\bfx) \equiv \bfB(\bfx,t_0)$, and hereafter take the magnetic field $\bfB(\bfx)$ and its Fourier $\tilde\bfB(\bfk)$ without an explicit time argument to be evaluated at $t_0$. We note that the scaling $\bfB(\bfx,t) = \bfB(\bfx)/a^2(t)$, of the magnetic field with scale factor, as been used in the derivation of Eq.~\eqref{eqn:EOM}.
We work in SI units, with $\mu_0$ the magnetic permeability of the vacuum, and the energy density in the magnetic field is $\rho_B=\bfB^2/(2\mu_0)$.\footnote{If Gaussian units are used, then $\mu_0 \to 4\pi$ and $\rho_B= \bfB^2/(8\pi)$.}
The fractional matter perturbation is then obtained as the special solution of this differential equation as
\begin{equation}
     \delta_m(\bfx,t) = M(t) \alpha v(\bfx).
\label{eqn:deltam}     
\end{equation}
where we define $\alpha\equiv f_b/\left(\mu_0 \bar\rho_{b,0}\right)$ to declutter.
The time dependence $M(t)$ satisfies,
\begin{equation}
     \ddot M(t) + 2 H(t) \dot M(t) - 4 \pi G \bar\rho_m (t) M(t) = \frac{1}{a^3(t)},
\end{equation}
with initial conditions $M(t_r)=\dot M(t_r)=0$ imposed at the time $t_r$ of recombination.  Under the approximation that the Universe is fully matter dominated at recombination, this becomes
\begin{equation}
     M(t) \to \frac{9}{10} t_r^2 \frac{D_+(t)}{D_+(t_r)},
\end{equation}
at $t\gg t_r$, where $D_+(t)$ is the linear-theory growth function ($\propto t^{2/3}$ during matter domination).  In practice, though, in current work the equation is solved numerically to take into account the fact that the Universe is not fully matter dominated at $t_r$. 

\section{The matter power spectrum}

The matter power spectrum $P(k,t)$ is defined as
\begin{equation}
    \left\langle \tilde \delta_m(\bfk,t) \tilde\delta_m^*(\bfk',t) \right\rangle = (2\pi)^3 \delta_D(\bfk- \bfk') P(k,t),
\end{equation}
where $\delta_D(\bfk)$ is the Dirac delta function,
\begin{equation}
    \tilde \delta_m(\bfk,t) = \int d^3x \, e^{-i \bfk \cdot \bfx} \delta_m(\bfx,t),
\end{equation}
is the Fourier transform of the density field, and
\begin{equation}
     \delta_m(\bfx,t) = \int \frac{d^3k}{(2\pi)^3} \tilde\delta_m(\bfk,t) e^{ i \bfk \cdot \bfx},
\end{equation}     
is the inverse Fourier transform.

Given Eq.~\eqref{eqn:deltam}, the total matter power spectrum can be written,
\begin{equation}
     P(k,t) = D_+^2(t) P(k,t_0) + M^2(t) \Pi(k),
\end{equation}
where $M^2(t) \Pi(k)$ is the magnetic-field-induced matter power spectrum.

\section{The mode-coupling integral}

We now calculate $\Pi(k)$ assuming some spectrum of PMFs.
We define the power spectrum $P_B(k)$ of the magnetic field (today) by,
\begin{equation}
     \left\langle{ B_p(\bfk) B_{p'}^*(\bfk')} \right\rangle
     = (2\pi)^3 \delta_D(\bfk-\bfk') \delta_{pp'} \frac12 P_B(k),
\label{eqn:PBeqn}
\end{equation}
where $B_p(\bfk)$ are coefficients in the Fourier expansion,
\begin{equation}
     \bfB(\bfx) = \sum_p \int \frac{d^3k}{(2\pi)^3} e^{i \bfk
     \cdot \bfx} B_p(\bfk) \hateps_p(\bfk),
\end{equation}
of the magnetic field today.
The two polarization (unit) vectors $\hateps_p(\bfk)$, $p=\{1,2\}$, are orthogonal to $\bfk$ and to each other.  Given the completeness relation,
\begin{equation}
     \sum_p \hateps_{p,\alpha}(\bfk) \hateps^*_{p,\beta}(\bfk) = \delta_{\alpha\beta} - \frac{k_\alpha k_\beta}{k^2},
\end{equation}
for the polarization vectors (where the subscripts $\alpha,\beta$ denote Cartesian components), the Cartesian components $B_\alpha(\bfk)$ of the magnetic field satisfy,
\begin{equation}
     \left\langle{ B_\alpha(\bfk) B_\beta^*(\bfk')} \right\rangle
     = (2\pi)^3 \delta_D(\bfk-\bfk') \frac12 \left( \delta_{\alpha\beta} - \frac{k_\alpha k_\beta}{k^2} \right) P_B(k),
\end{equation}
as in prior literature.

From Eq.~\eqref{eqn:deltam}, it follows that \begin{equation}
     \Pi(k) =  \alpha^2 P_v(k),
\end{equation}
where $P_v(k)$ is the power spectrum for $v(\bfx)$,
defined by
\begin{equation}
     \left\langle \tilde v(\bfk) \tilde v^*(\bfk') \right\rangle = (2\pi)^3
     \delta_D(\bfk-\bfk') P_v(k),
\label{eqn:Pvdefinition}     
\end{equation}

\begin{figure}[t!]
    \centering
    \includegraphics[width=\textwidth/3]{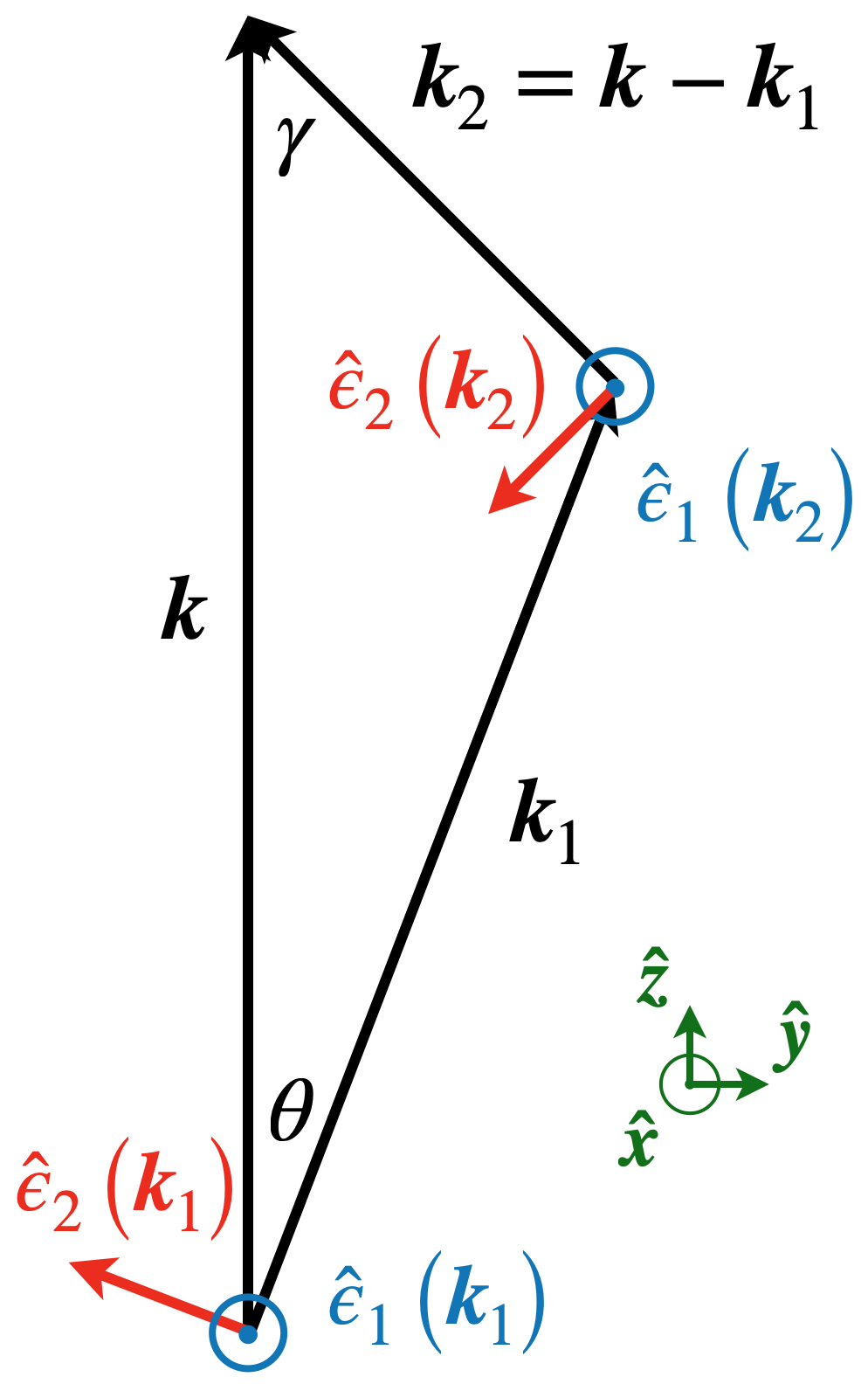}
    \caption{Schematic triangles of the vectors $\mathbf{k}$, $\mathbf{k}_1$, and $\mathbf{k}_2$ along with the polarization vectors $\hat{\epsilon}_{p_i}(\mathbf{k_i})$ orthogonal to $\mathbf{k_i}$ used to describe the magnetic field vectors. Only the terms in which $\hat{\epsilon}_{p_i}(\mathbf{k_i})$ both point in the $+\mathbf{\hat{x}}$ direction (out of the page) or in the $\mathbf{\hat{y}}-\mathbf{\hat{z}}$ plane (in the plane of the page) will contribute a nonzero integrand in Eq.~\eqref{eqn:Bfourier}.}
  \label{fig:triangle}
\end{figure}

The Fourier coefficients for $v(\bfx)$ are written
\begin{eqnarray}
     \tilde v(\bfk) &=& \sum_{p_1 p_2}\int \frac{d^3 k_1}{(2\pi)^3}\int
     \frac{d^3k_2}{(2\pi)^3} (2\pi)^3 \delta_D(\bfk_1+\bfk_2-\bfk) \nonumber \\
     & & \bfk \cdot \left[
     \left( \bfk_1 \times \hateps_{p_1}(\bfk_1) \right) \times
     \hateps_{p_2}(\bfk_2) \right] B_{p_1}(\bfk_1) B_{p_2}(\bfk_2),\nonumber \\
\label{eqn:Bfourier}     
\end{eqnarray}
where $p_i$ is the index of the polarization vectors identified with the vector ${\bfk}_i$.
The Dirac delta function restricts the three vectors $\bfk$, $\bfk_1$, and $\bfk_2$ in
Eq.~\eqref{eqn:Bfourier} to be the sides of a triangle, as shown in Fig.~\ref{fig:triangle}.  We take
$\bfk \parallel {\bf \hat z}$, and choose the triangle to be in the
$y$-$z$ plane.  One of the polarization unit vectors for $\bfk_1$ can be
chosen to be in the ${\bf \hat x}$
direction and the other in the plane, and similarly for
$\bfk_2$.  There are four possible combinations of the
polarization vectors for $\bfk_1$ and $\bfk_2$.  However, only two of these give nonzero ${\bf \hat k} \cdot \left[({\bf \hat k}_1
\times \hateps_{p_1}(\bfk_1))\times \hateps_{p_2}(\bfk_2) \right]$; the combinations where
${\bf \hat k}_1 \times \hateps_{p_1}(\bfk_1)$ and
$\hateps_{p_2}(\bfk_2)$ are either both out of the plane or
both in the plane contribute nothing.  A nonzero contribution
arises only when one is in the plane and the other is out. As
shown in Fig.~\ref{fig:triangle}, one of these combinations
gives
\begin{equation}
    {\bf \hat k}\cdot \left[({\bf \hat k}_1 \times \hateps_{p_1}(\bfk_1))\times \hateps_{p_2}(\bfk_2) \right] = \cos\theta = {\bf \hat k} \cdot {\bf \hat k}_1 \equiv \mu_1,
\end{equation}
and the other gives
\begin{equation}
    {\bf \hat k} \cdot \left[ ({\bf \hat k}_1 \times \hateps_{p_1}(\bfk_1))\times \hateps_{p_2}(\bfk_2) \right]=  \cos\gamma ={\bf \hat k} \cdot {\bf \hat k}_2 \equiv \mu_2.
\end{equation}
The needed Fourier amplitudes can thus be written, 
\begin{eqnarray} 
     \tilde v(\bfk) &=& \int \frac{d^3k_1}{(2\pi)^3}\int
     \frac{d^3k_1}{(2\pi)^3} (2\pi)^3 \delta_D(\bfk-\bfk_1-\bfk_2) k k_1 \nonumber \\
     & & \left[ B_1(\bfk_1) B_1(\bfk_2) ({{\bf \hat k}}
     \cdot {{\bf \hat k}}_1) + B_2(\bfk_1) B_2(\bfk_2) ({{\bf \hat k}}
     \cdot {\bf {\hat k}}_2) \right],\nonumber\\
\label{eqn:Bfouriertwo}
\end{eqnarray}
where the subscripts $p$ on $B_p(\bfk)$ refer to the two polarization amplitudes associated with $\bfk$.

Assuming a Gaussian distribution for the magnetic fields, we use Wick's theorem along with Eq.~\eqref{eqn:Pvdefinition}, and find the power spectrum to be
\begin{eqnarray}
     P_v(k) &=& \frac14 k^2 \int \frac{d^3k_1}{(2\pi)^3} \int
     \frac{d^3k_2}{(2\pi)^3} (2\pi)^3 \delta_D(\bfk-\bfk_1-\bfk_2)
     \nonumber \\
     & & \times P_B(k_1) P_B(k_2) \left[ 
     k_1^2 (\mu_1^2+\mu_2^2) + 2 k_1 k_2 \mu_1 \mu_2 \right]. \nonumber \\
\label{eqn:symmetric}     
\end{eqnarray}
We then note that $\bfk_1$ and $\bfk_2$ are here dummy
variables that are integrated over, and so we can make the replacement $k_1^2 \to k_2^2$ in the first term.  We then use $k= k_1 \mu_1+k_2\mu_2$ and $k_2^2= k^2+k_1^2 -2 k k_1 \mu_1$ (as will be imposed by the Dirac delta function) and $d^3k_1 = (2\pi) k_1^2\, dk_1\, d\mu_1$ and add the prefactor relating $\Pi(k)$ to $P_v(k)$ to obtain,
\begin{eqnarray}
     \Pi(k) &=& \left(\frac{\alpha k}{4\pi}\right)^2
     \int_0^\infty k_1^2\, dk_1
     \int_{-1}^1 d\mu\, \nonumber \\
     & &  \times  P_B(k_1) P_B\left(\sqrt{k^2+k_1^2 -2k k_1
     \mu} \right)\nonumber \\
     & &  \times \left[k^2+(k^2-2 k k_1\mu)\mu^2 \right],
\label{eqn:ourPi}     
\end{eqnarray}
where we dropped the subscript $1$, setting $\mu=\mu_1$.

If we proceed as above but without making the replacement $k_1\to k_2$ in the integrand in Eq.~\eqref{eqn:symmetric}, then we arrive at the same result,
\begin{eqnarray}
     \Pi(k) &=&  \left(\frac{\alpha}{4\pi}\right)^2
     k^3  \int_0^\infty k_1^3\, dk_1
     \int_{-1}^1 d\mu\, \nonumber \\
     & &  \times \frac{P_B(k_1) P_B\left(\sqrt{k^2+k_1^2 -2k k_1
     \mu} \right)}{k^2+k_1^2 -2k k_1}\nonumber \\
     & &  \times \left[2k^2 \mu +k k_1 (1-5\mu^2)
     + 2k_1^2 \mu^3 \right],
\label{eqn:priorPi}     
\end{eqnarray}
for the power spectrum derived in Ref.~\cite{kim96}.  A simple numerical integration verifies that Eqs.~\eqref{eqn:ourPi} and \eqref{eqn:priorPi} agree.  The simpler expression, Eq.~\eqref{eqn:ourPi}, avoids a quantity that goes to zero in the denominator of the integrand and is thus a bit more easily evaluated numerically.

\section{Discussion}

\begin{figure}[h!]
    \centering
    \includegraphics[width=\columnwidth]{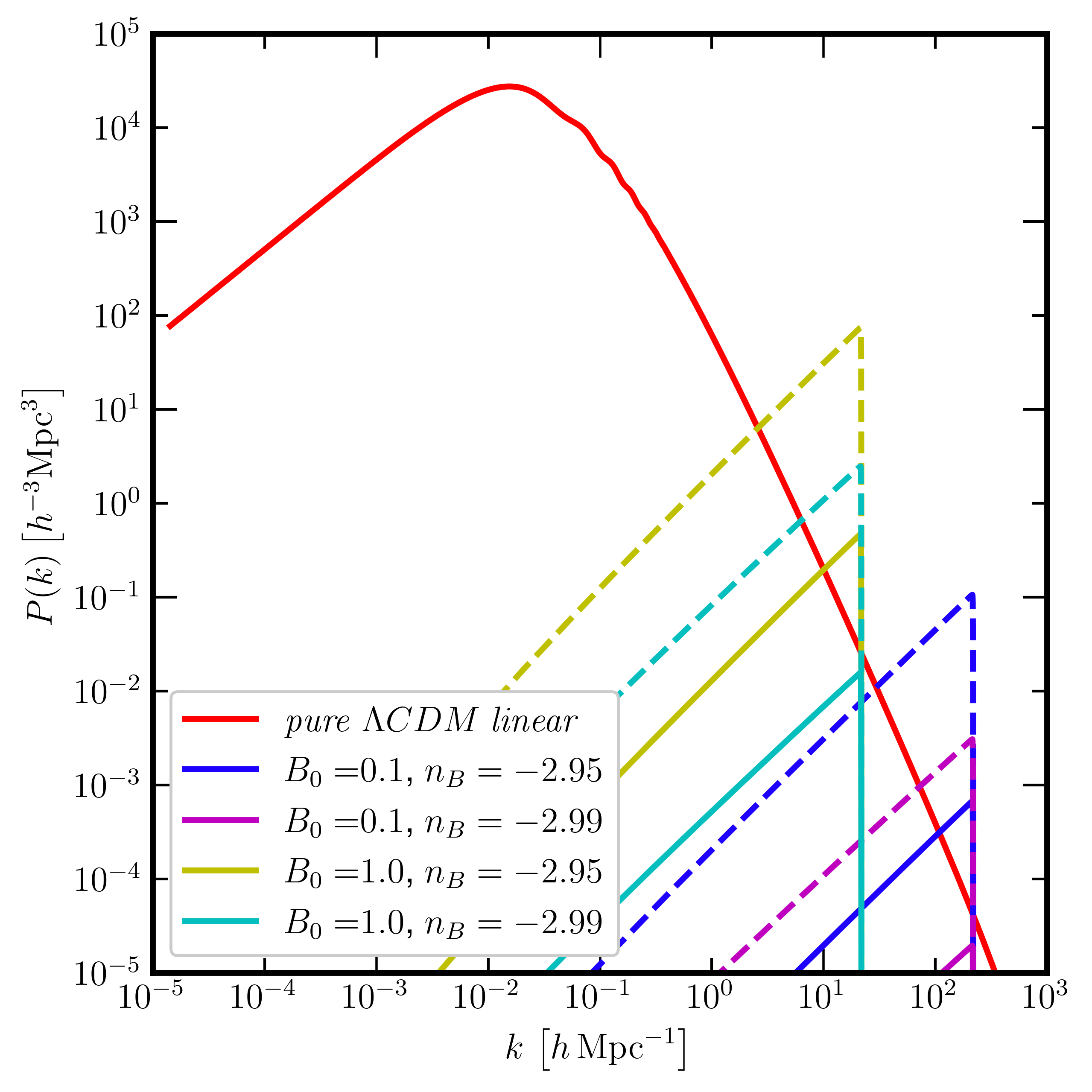}
    \caption{A comparison plot for the magnetic contribution to the linear matter power spectrum\footnote{The linear power spectrum was computed using \href{https://github.com/lesgourg/class_public}{\texttt{CLASS}}~\cite{Blas:2011rf}.} for $P_B(k)=A_Bk^{n_B}$. The dashed lines correspond to the contribution without the  $1/(4\pi)^2$ factor, reproduced according to Refs.~\cite{pandey15, subramanian16}. Whereas the solid lines correspond to the contribution according to the derivation in this work. The parameter $B_0$ represents the magnetic field strength today in nGauss, and the cutoffs are at the magnetic Jeans scale; both are determined in a similar manner as described in Refs.~\cite{pandey15, subramanian16}.}
    \label{fig:lin_ps}
\end{figure}

Our analytic result for the matter power spectrum is smaller by a factor of $(4\pi)^2$ than the analytic expressions in several prior papers \cite{gopal03,tashiro06a,tashiro06b,fedeli12} (and an earlier version of Ref.~\cite{Adi:2023qdf}), as we visualize in Fig.~\ref{fig:lin_ps}.  We have not been able to trace the origin of these discrepancies, but in at least one case \cite{gopal03} we believe it is due partially to inconsistencies in Fourier conventions.  We have also checked that the errors are propagated in the numerical work and also in other papers \cite{pandey13,pandey15,Sanati:2020oay,Sethi:2004pe,Pandey:2012kk} that use the results.

We advocate using Eq.~\eqref{eqn:ourPi} in future work on density perturbations from PMFs.

\begin{acknowledgments}

We thank Ely Kovetz for suggestions.  TA is supported by a Negev PhD fellowship awarded by the BGU Kreitmann School.  HAC is supported by the National Science Foundation Graduate Research Fellowship under Grant No.\ DGE2139757.  MK is supported by NSF Grant No.\ 1818899 and the Simons Foundation.
\end{acknowledgments}

%\clearpage

\bibliography{refs}

\end{document}